\documentclass[prb,preprint,superscriptaddress,groupedaddress,showpacs,preprintnumbers,amsmath,amssymb]{revtex4}
\usepackage{graphicx}
\usepackage{dcolumn}
\usepackage{bm}

\begin{document}


\title{Spin tunneling through an indirect barrier}


\author{Titus Sandu}
 \email{titus.sandu@umontreal.ca}
\affiliation{D\'{e}partement de Chimie, Universit\'{e} de Montr\'{e}al, 
C.P. 6128, succursale Centre-ville, Montr\'{e}al, Qu\'{e}bec H3C 3J7, Canada
 }%
  
\author{Athanasios Chantis}%
\affiliation{Department of Chemical and Materials Engineering,
Arizona State University, Tempe, AZ, 85287
\\}%

\author{Radu Iftimie}%
\affiliation{D\'{e}partement de Chimie, Universit\'{e} de Montr\'{e}al, 
C.P. 6128, succursale Centre-ville, Montr\'{e}al, Qu\'{e}bec H3C 3J7, Canada
 }%



\date{\today}

\begin{abstract}
Spin-dependent tunneling through an indirect bandgap barrier like the 
GaAs/AlAs/GaAs heterostructure along [001] direction is 
studied by the tight-binding method. The tunneling is characterized by the proportionality of 
the Dresselhaus Hamiltonians at $\Gamma $ and  $X$ points in the barrier and by Fano resonances 
(i.e. pairs of resonances and anti-resonances or zeroes in transmission).
The present results suggest that large spin polarization can be obtained 
for energy windows that exceed significantly the spin 
splitting. The width of these energy windows are mainly determined by the energy difference between the resonance and its associated zero, 
which in turn, increases with the decrease of barrier transmissibility at direct tunneling. 

We formulate two conditions that are necessary for the existence of energy 
windows with large polarization: First, the resonances must be well
separated such that their corresponding zeroes are not pushed away from the real axis by mutual interaction.
Second, the relative energy order of the resonances in the two spin channels must be the same as
the order of their corresponding zeroes.  

The degree to which the first condition is fulfilled is determined by
the barrier width and the longitudinal effective mass at $X$ point. In contrast, 
the second condition can be satisfied by choosing an appropriate 
combination of spin splitting strength at $X$ point and transmissibility through the direct 
barrier.

\end{abstract}


\maketitle


\section{Introduction}

The spin rather than the charge of carriers 
has attracted a lot of interest leading to a new field of electronics dubbed 
spintronics.\cite{Wolf01,Zutic04} In this context, spin-polarized transport in 
non-magnetic semiconductor structures and spin-dependent 
properties originating from the spin-orbit interaction are a promising road to 
spin based devices. \cite{Awschalom02}  

Despite the progress that has been made, \cite{Rashba00,Hanbicki02} spin injection from ferromagnetic 
leads proved to be very challenging. \cite{Schmidt00} Consequently,
spin-dependent transport in nanostructures comprised of non-magnetic semiconductors has  been
the focus of extensive work in the past years.\cite{Awschalom02} Recent theoretical research has 
suggested that the current 
resulting from electron tunneling through zinc-blende
semiconductor single\cite{Perel03,Tarasenko04} or double barrier structures\cite{Glazov05} can be highly spin 
polarized. The origin of the spin-dependent tunneling in these structures stems from the fact
that the barrier material lacks center of inversion. 

In the effective 
mass approximation, the electron Hamiltonian of a 
zinc-blende structure has an additional spin-dependent $k^3$ coupling called 
the Dresselhaus term\cite{Dresselhaus55}

\begin{equation}
\label{eq:DV}
H_D = \gamma \left[ {\sigma _x k_x \left( {k_y^2 - k_z^2 } \right) + \sigma 
_y k_y \left( {k_z^2 - k_x^2 } \right) + \sigma _z k_z \left( {k_x^2 - k_y^2 
} \right)} \right],
\end{equation}

\noindent
where $\sigma _i $ are the Pauli matrices, and $k_x $, $k_y $, and $k_z $ are the 
components of electron wave vector. For a barrier along $\left[ {001} 
\right]$ direction the Dresselhaus Hamiltonian is reduced to

\begin{equation}
\label{eq:DB}
H_D = \gamma \left( {\sigma _x k_x - \sigma _y k_y } \right)\frac{\partial 
^2}{\partial ^2z^2}.
\end{equation}

\noindent
Perel, Tarasenko, and coworkers\cite{Perel03,Tarasenko04} showed that the 
$\Gamma$ point Hamiltonian in Eq.~(\ref{eq:DB}) induces an effective mass 
correction leading to a spin-polarized transmission. It is important to emphasize
that  the spin 
dependent part of the effective mass Hamiltonian at $X$ point is \cite{Dresselhaus55,Yvchenko97}

\begin{equation}
\label{eq:DX}
H_D^X = \beta \left( {\sigma _x k_x - \sigma _y k_y } \right)
\end{equation}

\noindent
and therefore proportional to the Hamiltonian in Eq.~(\ref{eq:DB}). The Hamiltonians 
in (\ref{eq:DB}) and (\ref{eq:DX}) are diagonalized by

\begin{equation}
\label{eq:EV}
\chi _\pm = \frac{1}{\sqrt 2 }\left( {{\begin{array}{*{20}c}
 1 \hfill \\
 { \mp \,e^{ - i\varphi }} \hfill \\
\end{array} }} \right)
\end{equation}

\noindent
with $\varphi $ the polar angle of the wave vector $k_t$ in 
$xy$ plane. Therefore the spin states are not mixed by the 
interaction between $X$ states and $\Gamma$ states in the barrier. 
 
\begin{figure*}
\includegraphics{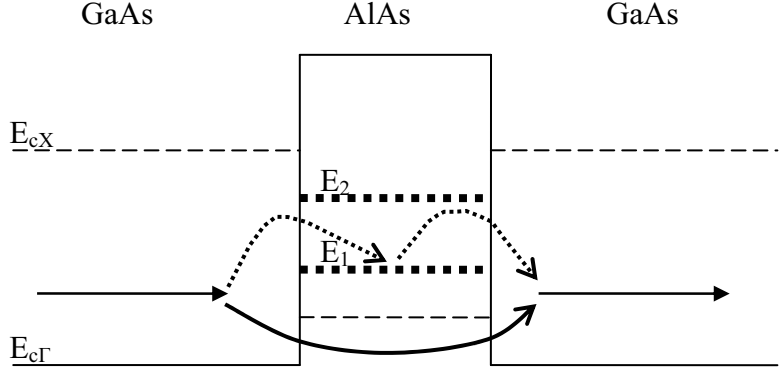}
\caption{\label{fig:1}  Tunneling through GaAs/AlAs/GaAs barrier. $\Gamma$ band edges are 
shown  by solid line and 
$X$ band edges are represented by dashed line. $E_{1}$ and $E_{2}$ are $X$-valley quasi-bound states in the AlAs 
barrier. Direct tunneling ($\Gamma$-$\Gamma$-$\Gamma $) is shown by solid curly arrow and 
$\Gamma$ -$X$-$\Gamma $ tunneling is depicted by dotted curly arrow.}
\end{figure*}

Spin-dependent transport can be studied using numerous 
treatments such as the $k \cdot p$ approach, full-band tight-binding calculations, and
ab initio methods. For various reasons the theoretical study of spin tunneling through an 
indirect barrier like GaAs/AlAs/GaAs has not been fully 
addressed before. $k \cdot p$ cannot fully address the problem because the AlAs 
barrier accommodates at least one quasi-bound state into $X$-valley. Thus, beside 
the $\Gamma$-$\Gamma$-$\Gamma $ tunneling, which occurs through the  
higher $\Gamma $-valley, one must also consider the tunneling ($\Gamma $-$X$-$\Gamma $) 
through the lower $X$-barrier (see Fig.~\ref{fig:1}). The $k \cdot p$ method is a perturbative 
method that can be ``tuned'' for the necessities of 
spin-dependent processes (for instance, see 
Ref.~\onlinecite{Rougemaille05}, in which spin-dependent evanescent states in the band gap are studied). 
In contrast, empirical tight-binding methods provide a treatment of the full 
Brillouin zone, but they lack the complete description of the Dresselhaus 
term when the spin-orbit is introduced.\cite{Boykin98} 
This is due to the fact that the orthogonality assumption in tight binding models
is incompatible with the formulation of the spin-orbit interaction. \cite{Sandu05}  
In principle, the above shortcomings should be overcome by utilizing \emph{ab-initio} 
density functional theory methods. However, these methods suffer on the side of 
bandgap reproducibility. \cite{Jones89}

Spin dependent tunneling has been recently analyzed with a 1-band envelope-function model. \cite{Mishra05} 
In their study,\cite{Mishra05} the authors neglected the ($\Gamma$-$\Gamma$-$\Gamma $) tunneling and the 
presence of $X$-valley quasi-bound states in the AlAs barrier. However, spin tunneling through the 
indirect barrier of the GaAs/AlAs/GaAs heterostructure 
shows another peculiar property. The confined $X$ states in the AlAs barrier
interact\cite{Beresford89} with the continuum $\Gamma $ states in GaAs
forming Fano resonances\cite{Fano61} (i.e.:pairs of resonances/anti-resonances). 
In this paper we demonstrate that one can use the proximity in the resonance and anti-resonance 
states in conjunction with the spin 
splitting produced by the spin-dependent Hamiltonian to obtain a large 
degree of spin polarization within the range between the resonance and anti-resonance 
energy. For this purpose we devise a spin-dependent tight-binding model that provides 
a realistic view of the spin-dependent tunneling through an indirect 
barrier. We convert the spin-dependent effective mass Hamiltonians for a 
single band to their tight-binding versions following the recipes of 
Ref.~\onlinecite{Frensley94}. The coupling between $\Gamma $ and X valleys is made 
according to Ref.~\onlinecite{Fu93}. 

The paper is organized as follows. In next section a simple model is analyzed 
in order to gain insight into the physics of spin-dependent tunneling. The 
third section contains a realistic tight-binding model and the numerical 
results. Conclusions are drawn in the fourth section.

\section{Tight-binding model of spin-dependent tunneling}

Consider a simple tight-binding (TB) model of the 
spin-dependent tunneling through an indirect barrier. The main assumptions 
for this TB model are: spin states are degenerate in left/right lead (bulk-like states) and the 
Dresselhaus Hamiltonians are proportional for $\Gamma $ and X states in the 
barrier, so we can assume that the spin states in the leads are  
eigenvalues of the Dresselhaus Hamiltonian. 
\paragraph{Spin tunneling through an indirect barrier.} 
The Hamiltonian of the 
system is

\begin{widetext}
\begin{equation}
\label{eq:1}
\begin{array}{l}
 H = \sum\limits_{n = - \infty ,\sigma = \uparrow , \downarrow }^{ - 1} 
{\left[ {\varepsilon {\kern 1pt} c_{n,\sigma }^ + c_{n,\sigma } + \left( 
{t\,c_{n - 1,\sigma }^ + c_{n,\sigma } + h.c.} \right)} \right]} +\\  
 \quad \quad \sum\limits_{n = 1,\sigma = \uparrow , \downarrow }^\infty 
{\left[ {\varepsilon {\kern 1pt} c_{n,\sigma }^ + c_{n,\sigma } + \left( 
{t\,c_{n,\sigma }^ + c_{n + 1,\sigma } + h.c.} \right)} \right]} + \\ 
 \quad \quad \varepsilon _1 c_{0, \uparrow }^ + c_{0, \uparrow } + 
\varepsilon _2 c_{0, \downarrow }^ + c_{0, \downarrow } + \left( {V_1 c_{0, 
\uparrow }^ + c_{ - 1, \uparrow } + V_2 c_{0, \downarrow }^ + c_{ - 1, 
\downarrow } + h.c.} \right) + \\
 \quad \quad \left( {V_1 c_{0, \uparrow }^ + c_{1, \uparrow } + V_2 c_{0, 
\downarrow }^ + c_{1, \downarrow } + h.c.} \right) + \left( {t_1 c_{ - 1, 
\uparrow }^ + c_{1, \uparrow } + t_2 c_{ - 1, \downarrow }^ + c_{1, 
\downarrow } + h.c.} \right) \\ 
 \end{array} .
\end{equation}
\end{widetext}

\noindent
The first two terms on the right hand side of Eq.~(\ref{eq:1}) are the Hamiltonians of the 
contacts (leads), where $\varepsilon $ and $t$ are the on-site energy and transfer integral, respectively (spin degenerate ). 
$c_{n\sigma }^ + / c_{n\sigma } $ 
is the creation/annihilation of an electron with spin $\sigma $ on site $n$. The remaining part is the Hamiltonian of the barrier and its 
coupling to the leads. The active region is modeled by three sites: 
site $ - 1$ that is like the left hand side contact, site $0$ that is the 
actual barrier, and site $1$ that is like the right hand side contact. Thus 
the effective left/right hand side contact ends/starts at site $ - 2 / 2$. The 
matrix form of the Hamiltonian for the sites -1, 0, and 1 (in fact $E - 
H$, where $E$ is energy) with appropriate boundary conditions for an open 
system is\cite{Datta95}

\begin{widetext}
\begin{equation}
\label{eq:2}
E - H = {\begin{array}{*{20}c}
 \hfill & { - 1 \uparrow } \hfill & { - 1 \downarrow } \hfill & {0 \uparrow 
} \hfill & {0 \downarrow } \hfill & {1 \uparrow } \hfill & {1 \downarrow } 
\hfill \\
 { - 1 \uparrow } \hfill & {E - \varepsilon - \Sigma _L \left( E \right)} 
\hfill & 0 \hfill & {-V_1 } \hfill & 0 \hfill & {-t_1 } \hfill & 0 \hfill \\
 { - 1 \downarrow } \hfill & 0 \hfill & {E - \varepsilon - \Sigma _L \left( 
E \right)} \hfill & 0 \hfill & {-V_2 } \hfill & 0 \hfill & {-t_2 } \hfill \\
 {\quad 0 \uparrow } \hfill & {-V_1^\ast } \hfill & 0 \hfill & {E - 
\varepsilon _1 } \hfill & 0 \hfill & {-V_1^\ast } \hfill & 0 \hfill \\
 {\quad 0 \downarrow } \hfill & 0 \hfill & {-V_2^\ast } \hfill & 0 \hfill & 
{E - \varepsilon _2 } \hfill & 0 \hfill & {-V_2^\ast } \hfill \\
 {\quad 1 \uparrow } \hfill & {-t_1^\ast } \hfill & 0 \hfill & {-V_1 } \hfill 
& 0 \hfill & {E - \varepsilon - \Sigma _R \left( E \right)} \hfill & 0 
\hfill \\
 {\quad 1 \downarrow } \hfill & 0 \hfill & {-t_2^\ast } \hfill & 0 \hfill & 
{-V_2 } \hfill & 0 \hfill & {E - \varepsilon - \Sigma _R \left( E \right)} 
\hfill \\
\end{array} } 
\end{equation}
\end{widetext}

\noindent
where $\Sigma _{L,R} \left( E \right)$ are the self-energies of the 
semi-infinite parts, i.e.,

\begin{equation}
\label{eq:3}
\Sigma _{L,R} \left( E \right) = t^ + \frac{1}{E - H_{L,R} + i\delta }\,\,t ,
\end{equation}

\noindent
where $H_{L,R} $ are the Hamiltonians of the semi-infinite left and right hand 
sides and $\delta$ is an infinitesimal positive number. The retarded Green function

\begin{equation}
\label{eq:4}
G_{L,R}^R \left( E \right) = \frac{1}{E - H_{L,R} + i\delta }
\end{equation}

\noindent
of the left/right hand side semi-infinite contact in Eq.~(\ref{eq:3}) is actually the 
diagonal part $G_L^R \left( E \right)_{ - 2, - 2} / G_R^R \left( E 
\right)_{2,2} $ representing the sites $ - 2$/2. The expressions of these Green function elements 
can be found from 
their equation of motion and the use of the finite difference equation method.\cite{Chen89}  
If we consider the 
parameterization $\varepsilon - E = 2{\kern 1pt} {\kern 1pt} t\,\cos \left( 
{ka} \right)$, with $k$ a complex parameter and $a$ a lattice constant 
parameter, one obtains the following equation for self-energies,

\begin{equation}
\label{eq:13}
\Sigma _{L,R} = - t\,\,e^{ika} .
\end{equation}

\noindent
Therefore, the Green function $G^R$ for the sites -1, 0, and 1 with the 
boundary conditions for an open system is

\begin{widetext}
\begin{equation}
\label{eq:14}
G^R\left( E \right) = \left[ {{\begin{array}{*{20}c}
 { - t\,e^{ - ika}} \hfill & 0 \hfill & {-V_1 } \hfill & 0 \hfill & {- t_1 } 
\hfill & 0 \hfill \\
 0 \hfill & { - t\,e^{ - ika}} \hfill & 0 \hfill & {- V_2 } \hfill & 0 \hfill 
& {- t_2 } \hfill \\
 {-V_1^\ast } \hfill & 0 \hfill & {E - \varepsilon _1 } \hfill & 0 \hfill & 
{-V_1^\ast } \hfill & 0 \hfill \\
 0 \hfill & {- V_2^\ast } \hfill & 0 \hfill & {E - \varepsilon _2 } \hfill & 0 
\hfill & {- V_2^\ast } \hfill \\
 {- t_1^\ast } \hfill & 0 \hfill & {-V_1 } \hfill & 0 \hfill & { - t\,e^{ - 
ika}} \hfill & 0 \hfill \\
 0 \hfill & {- t_2^\ast } \hfill & 0 \hfill & {- V_2 } \hfill & 0 \hfill & { - 
t\,e^{ - ika}} \hfill \\
\end{array} }} \right]^{ - 1} .
\end{equation}
\end{widetext}

\noindent
We notice that the Hamiltonian is not hermitian due to open boundary 
conditions. To calculate the transmission probability from site $1$ to $N$ 
we use the formula\cite{Datta95}

\begin{equation}
\label{eq:32}
T\left( {E,k_t } \right) = \Gamma _L \left( {E,k_t } \right)\left| 
{G_{1,N}^R \left( {E,k_t } \right)} \right|^2\Gamma _R \left( {E,k_t } 
\right),
\end{equation}

\noindent
with

\begin{equation}
\label{eq:33}
\Gamma _{L,R} \left( {E,k_t } \right) = i\left[ {\Sigma _{L,R} \left( {E,k_t 
} \right) - \Sigma _{L,R}^\ast \left( {E,k_t } \right)} \right].
\end{equation}

\noindent
Since the Dresselhaus Hamiltonians are proportional at $\Gamma $ 
and X points in the barrier\cite{Dresselhaus55,Yvchenko97} we can solve separately for each spin. 
The Green function for 'spin up' is

\begin{equation}
\label{eq:15}
G_ \uparrow ^R \left( E \right) = \left[ {{\begin{array}{*{20}c}
 { - t\,e^{ - ika}} \hfill & {-V_1 } \hfill & {- t_1 } \hfill \\
 {-V_1^\ast } \hfill & {E - \varepsilon _1 } \hfill & {-V_1^\ast } \hfill \\
 {- t_1^\ast } \hfill & {-V_1 } \hfill & { - t\,e^{ - ika}} \hfill \\
\end{array} }} \right]^{ - 1} .
\end{equation}

\noindent
A similar equation is obtained for the 'spin down'. The poles of $G^R$ are the 
solutions of the determinant equation

\begin{equation}
\label{eq:16}
\Delta = \left| {{\begin{array}{*{20}c}
 { - t\,e^{ - ika}} \hfill & {-V_1 } \hfill & {- t_1 } \hfill \\
 {-V_1^\ast } \hfill & {E - \varepsilon _1 } \hfill & {-V_1^\ast } \hfill \\
 {- t_1^\ast } \hfill & {-V_1 } \hfill & { - t\,e^{ - ika}} \hfill \\
\end{array} }} \right| = 0,
\end{equation}

\noindent
while the zeroes of the transmission are the zeroes of the Green function 
$\left( {G_ \uparrow ^R } \right)_{1,3} $ relating the sites -1 and 1,

\begin{equation}
\label{eq:17}
\left( {G_ \uparrow ^R } \right)_{1,3} \left( E \right) = \frac{\left| 
{{\begin{array}{*{20}c}
 {-V_1 } \hfill & {E - \varepsilon _1 } \hfill \\
 {- t_1 } \hfill & {-V_1^\ast } \hfill \\
\end{array} }} \right|}{\Delta } = 0 .
\end{equation}

\noindent
The equation for poles reads

\begin{equation}
\label{eq:18}
\left( {E - \varepsilon _1 } \right)\left[ {\left| {t_1 } \right|^2 - 
t^2\,e^{ - 2i\,ka}} \right] = \left| {V_1 } \right|^2\left[ {-t_1-t_{1}^{*} + 
2\,t\,e^{ - i\,ka}} \right]
\end{equation}

\noindent
and for zeroes

\begin{equation}
\label{eq:19}
\left( {E - \varepsilon _1 } \right)\,t_1 = -\left| {V_1 } \right|^2.
\end{equation}

\noindent
Since $t$ is basically the conduction band bandwidth and $t_{1}$ and $V_{1}$ are 
tunneling rates, then $V_1 ,t_1 < < t$, such that the pole is given by

\begin{equation}
\label{eq:20}
E \approx \varepsilon _1 - 2\frac{\left| {V_1 } \right|^2}{t}e^{ika}.
\end{equation}

\noindent
The equation for the zero is

\begin{equation}
\label{eq:21}
E = \varepsilon _1 - \frac{\left| {V_1 } \right|^2}{t_1 }.
\end{equation}

\begin{table}
\caption{\label{tab:table1} Matrix elements in eV of the nearest neighbor model outlined  in Eq. ~(\ref{eq:1}). }
\begin{ruledtabular}
\begin{tabular}{cccccccc}
& 
Indirect barrier  \par & 
RTD-like structure \par  \\
\hline
$t$& 
1.0& 
1.0 \\
$V_1$ & 
0.05 & 
0.005 \\
$V_2$ & 
0.052 & 
0.0052 \\
$t_1$ & 
0.05 & 
0.0 \\
$t_2$ & 
0.052 & 
0.0 \\
$\varepsilon _1$ & 
0.17 & 
0.17 \\
$\varepsilon _2$ & 
0.175 & 
0.175 \\

\end{tabular}
\end{ruledtabular}
\label{tab1}
\end{table}

\begin{figure*}
\includegraphics{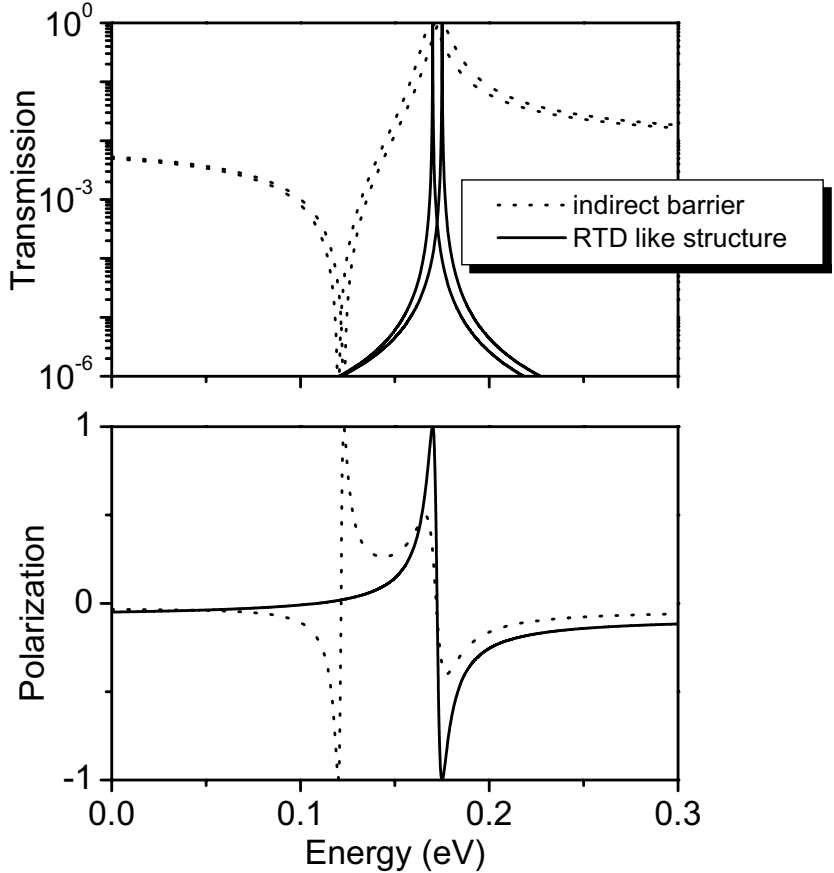}
\caption{\label{fig:2} Spin dependent transmission (upper panel) and spin polarization (lower panel) 
through an indirect barrier versus a resonant tunneling diode configuration. One can see 
the wide energy window of large polarization 
created by indirect tunneling. }
\end{figure*}

\noindent
Since we have $V_1 ,t_1 < < t$, 
the energy separation between the pole and the zero is about $V_{1}^{2}/t_1$. The resonances and zeroes occur at slightly different energies 
for the two spin channels, resulting in a large spin polarization 
due to the combination of a sharp increase in transmission at resonance followed by an abrupt 
decrease to zero at anti-resonance. 
Hence, the energy range of large spin polarization will depend on $V_1$ and $t_1 $ but not on 
the magnitude of spin splitting. 
One can notice that the energy separation between resonance and anti-resonance can be increased by decreasing 
$t_1$, i.e., increasing the width of the barrier. In Fig.~\ref{fig:2} we illustrate the above arguments with the 
parameters given in Table~\ref{tab:table1}. We also calculate the spin polarization with the equation

\begin{equation}
\label{eq:polarization}
P = \frac{T_ \uparrow - T_ \downarrow }{T_ \uparrow + T_ \downarrow } .
\end{equation}

\noindent
For comparison we also plot the case of the resonance 
tunneling diode (RTD) 
configuration. The RTD configuration is made by setting $t_1$ and $t_2$ to 0. To separate the spin resonances, 
we also have chosen 
the values of $V_1$ and $V_2$ ten times smaller 
than their values in the indirect-barrier configuration. Fig.~\ref{fig:2} shows that the indirect barrier 
configuration has a clear advantage over the RTD configuration.

\paragraph{Spin tunneling through a direct barrier.} 

Following the same procedure one can also calculate spin transmission 
through a direct barrier. The Green function of the open system is

\begin{equation}
\label{eq:22}
G^R\left( E \right) = \left[ {{\begin{array}{*{20}c}
 { - t\,e^{ - ika}} \hfill & 0 \hfill & {- t_1 } \hfill & 0 \hfill \\
 0 \hfill & { - t\,e^{ - ika}} \hfill & 0 \hfill & {- t_2 } \hfill \\
 {- t_1^\ast } \hfill & 0 \hfill & { - t\,e^{ - ika}} \hfill & 0 \hfill \\
 0 \hfill & {- t_1^\ast } \hfill & 0 \hfill & { - t\,e^{ - ika}} \hfill \\
\end{array} }} \right]^{ - 1}.
\end{equation}

\noindent
The Green function for the 'spin up' is

\begin{equation}
\label{eq:23}
G_ \uparrow ^R \left( E \right) = \left[ {{\begin{array}{*{20}c}
 { - t\,e^{ - ika}} \hfill & {- t_1 } \hfill \\
 {- t_1^\ast } \hfill & { - t\,e^{ - ika}} \hfill \\
\end{array} }} \right]^{ - 1}.
\end{equation}

\noindent
Using Eqs.~(\ref{eq:32}) and (\ref{eq:33}) we calculate the spin-dependent transmission 
probabilities for a direct barrier. The transmission probabilities are approximated by 

\begin{equation}
\label{eq:24}
T_ \uparrow \approx \frac{4 \left|  {t_1 } \right|^2}{t^2}\sin^2{ka}
\end{equation}

\noindent
and

\begin{equation}
\label{eq:25}
T_ \downarrow \approx \frac{4 \left| {t_2 } \right|^2}{t^2}\sin^2{ka},
\end{equation}

\noindent
i.e., the transmission  for 'spin up' is different from transmission for 'spin down' leading to spin 
polarization.

\section{Results of a realistic tight-binding model and Discussions}

In order to make quantitative assessments of the spin tunneling through an 
indirect barrier, we devise a spin-dependent tight-binding model for a system 
that is partitioned into layers. The layers from $ - \infty $ to $0$ and 
from $N + 1$ to $\infty $ are the contacts, while the layers from $1$ to $N$ 
are the active layers. First, we consider the spin independent Hamiltonian. 
The coupling between $\Gamma $ and $X$ states is treated similarly to Ref.~\onlinecite{Fu93}, 
with the Hamiltonian

\begin{equation}
\label{eq:27}
H = \left[ {{\begin{array}{*{20}c}
 {H_\Gamma } \hfill & {H_{\Gamma X} } \hfill \\
 {H_{X\Gamma } } \hfill & {H_X } \hfill \\
\end{array} }} \right].
\end{equation}

\noindent
$H_\Gamma $ and $H_X $ are the Hamiltonians at $\Gamma$ and $X$ point, respectively. $H_{\Gamma{X}}$ and 
$H_{X\Gamma}$ are the couplings between $\Gamma$ and $X$ at the interface layers. For simplicity, we do 
not distinguish between $X_1 $ and $X_3 $, such that 
$H_\Gamma $ and $H_X $ are single-band effective mass Hamiltonians that are converted to TB 
Hamiltonians according to the parameterization given in Ref.~\onlinecite{Frensley94}. This TB parameterization 
has been successfully used in quantum transport for non-equilibrium 
conditions and incoherent scattering processes.\cite{Lake97} The parameterization\cite{Frensley94,Lake97} 
is made for the effective mass 
Hamiltonian

\begin{equation}
\label{eq:28}
H_0 = \frac{ - \hbar ^2}{2}\frac{d}{d{\kern 1pt} z}\frac{1}{m^\ast \left( z 
\right)}\frac{d}{d{\kern 1pt} z} + V_k \left( z \right) + \frac{\hbar 
^2k_t^2}{2{\kern 1pt} {\kern 1pt} m_L^\ast },
\end{equation}

\noindent
where $m_L^\ast $ is the effective mass in the left contact, the effective 
mass is considered $z$-dependent, and the spatial dependence of the transverse 
energy has been incorporated in the transverse momentum ($k_t$) dependent potential:

\begin{equation}
\label{eq:29}
V_k \left( z \right) = V\left( z \right) + \frac{\hbar ^2k_t^2}{2m_L^\ast 
}\left( {\frac{m_L^\ast }{m\left( z \right)} - 1} \right).
\end{equation}

The corresponding tight-binding parameters for the non-diagonal part are:

\begin{equation}
\label{eq:30}
t_{ij} = \frac{\hbar ^2}{\left( {m_i + m_j } \right){\kern 1pt} \Delta ^2}
\end{equation}

\noindent
and the diagonal part is

\begin{equation}
\label{eq:31}
D_i \left( k_t \right) = \frac{\hbar ^2}{2{\kern 1pt} {\kern 1pt} \Delta 
^2}\left( {\frac{1}{m^ - } + \frac{1}{m^ - }} \right) + V_i \left( k_t 
\right).
\end{equation}

\noindent
In Eqs.~(\ref{eq:30}) and  ~(\ref{eq:31}), $m_i $ is the effective mass at site $i$ on the mesh of 
spacing $\Delta$, $V_i \left( k_t 
\right)$ is the potential at site $i,$ which also includes the band offsets, 
$m^ - = \frac{m_{i - 1} + m_i }{2}$, and $m^ + = \frac{m_i + m_{i + 1} }{2}$.

\noindent
The spin dependent Hamiltonian is expressed in the basis spanned by spinors 
~(\ref{eq:EV}), such that the Hamiltonian is diagonal in this basis. At $\Gamma $ 
point the spin dependent part of the effective mass Hamiltonians is 
introduced through the corrections to the effective masses defined in Eq.~(\ref{eq:DB}) 
and the effective potential defined in Eq.~(\ref{eq:29}). The spin-dependent 
part at $X$ is expressed through different band offsets for the two 
spin projections as one can see from Eq.~(\ref{eq:DX}). 

\begin{table}
\caption{\label{tab:table2} Dresselhaus coefficients at $\Gamma$ and $X$ points for GaAs and AlAs, 
calculated with 
the GW method\cite{Faleev04,Chantis05} with spin-orbit coupling included. The units are the atomic units. }
\begin{ruledtabular}
\begin{tabular}{cccccccc}
& 
GaAs  \par & 
AlAs \par  \\
\hline
$\gamma$ & 
 2.1 & 
0.85 \\
$\beta$ & 
0.0074 & 
 0.00077 \\

\end{tabular}
\end{ruledtabular}
\label{tab1}
\end{table}

\begin{figure*}
\includegraphics{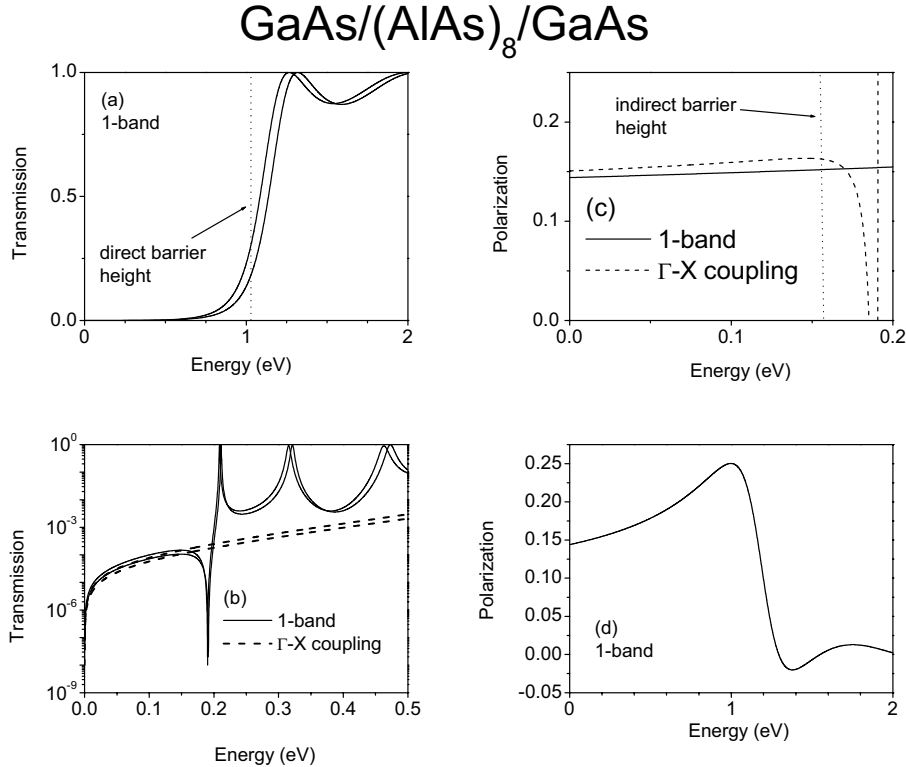}
\caption{\label{fig:3} (a) Spin dependent transmission probability of 1-band (direct) tunneling 
through GaAs/(AlAs)$_{8}$/GaAs barrier. (b) 
  Comparison between full (direct and indirect) and 1-band (direct) transmission probability  of 
a GaAs/(AlAs)$_{8}$/GaAs barrier. 
 (c) Comparison of spin polarization between full (direct and indirect) and 1-band (direct) 
electron transmission of a GaAs/(AlAs)$_{8}$/GaAs barrier. 
  (d) Spin polarization obtained from 1-band calculation over a broader energy range. }
\end{figure*}

\begin{figure*}
\includegraphics{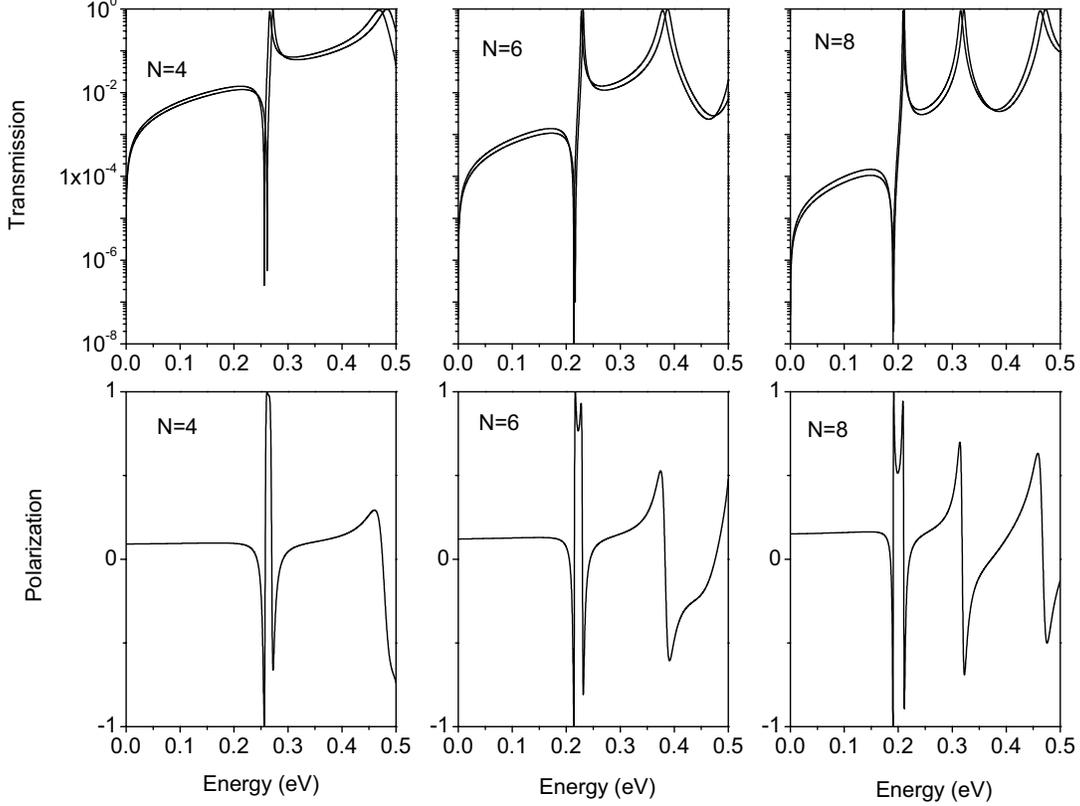}
\caption{\label{fig:4} Spin dependent transmission probability  and spin polarization of a 
GaAs/(AlAs)$_{N}$/GaAs heterostructure. 
$N$=4, 6, and 8. }
\end{figure*}

Our calculations are performed with the effective masses and 
band edges taken from Ref.~\onlinecite{Klimeck00}. The Dresselhaus parameters given in Table~\ref{tab:table2} 
are calculated with a quasi-particle self-consistent \emph{GW} (\emph{G}=Green function, 
\emph{W}=screened Coulomb interaction) method as in Ref.~\onlinecite{Faleev04,Chantis05} with spin-orbit 
coupling included perturbatively. 
The splitting at $\Gamma$ in GaAs found in the present work with the \emph{GW} method is three times smaller 
than the value used by Perel {\it et al.} in Ref. \onlinecite{Perel03}.\cite{Pikus}  The \emph{GW} spin splitting 
at $\Gamma$ for 
AlAs is about 2.5 times smaller than in GaAs. While for GaAs one can also use the experimental estimation 
for the splitting 
at the $\Gamma$ point \cite{Pikus} there are no such estimations for AlAs. Therefore to be consistent, 
we used the \emph{GW} values for both. 
The value of spin splitting at $X$ point is almost the same as the one obtained within the local 
density approximation (LDA)  with our 
FP-LMTO (full potential-linear muffin-tin orbitals) code. \cite{Methfessel00} The $X$-$\Gamma$ band offset for 
GaAs/AlAs heterostructure is chosen to be 160 meV.\cite{Mendez88}
Throughout the paper we have chosen a value $k_t = \frac{2\pi }{a}\times 
0.05$ for the transverse wave vector ($a$ is the lattice constant of GaAs).

In Fig.~\ref{fig:3} we compare the one-band model (direct tunneling) given by 
the effective 
mass Hamiltonian and the two-band model (direct and indirect tunneling) given by Eq.~(\ref{eq:27}). 
Below the indirect barrier, 
the main contribution to spin tunneling and polarization is provided by direct tunneling. However,
the tunneling and the polarization through $X$ states become dominant 
for energies above the indirect barrier. The result shows that the confinement of the $X$ 
states in the barrier increases the energy threshold at which 
the tunneling through $X$ states becomes dominant. In the calculations\cite{Mishra05} in which the confinement of the $X$ 
states in the barrier are neglected, the tunneling is predominantly indirect for energies 
slightly below the top of indirect barrier. Therefore, our calculations suggest that multi-band calculations 
are needed to fully describe the electron transport in these heterostructures.

In Fig.~\ref{fig:4} we show the 
transmission probability and spin polarization for GaAs/(AlAs)$_{N}$/GaAs heterostructures 
with $N$ = 4, 6, and 8. Energy 
windows with large 
polarization can be seen between the
resonance and its corresponding zero. The width of the window increases with the barrier width 
as it has been demonstrated in the previous section. Only the first resonance has a corresponding zero on 
the real axis, 
for the other resonances, the zeroes are pushed off the real axis.\cite{Bowen95} Therefore, if the 
resonances are close enough, no well 
defined window with large spin polarization can be found.  The possibility to obtain well separated 
resonances with zeroes 
on the real axis is controlled by the combination of the longitudinal effective mass in the barrier at $X$ 
point and the barrier width. 
A lighter longitudinal effective mass and/or a narrower barrier push farther apart the resonance 
energies in the barrier.

\begin{figure*}
\includegraphics{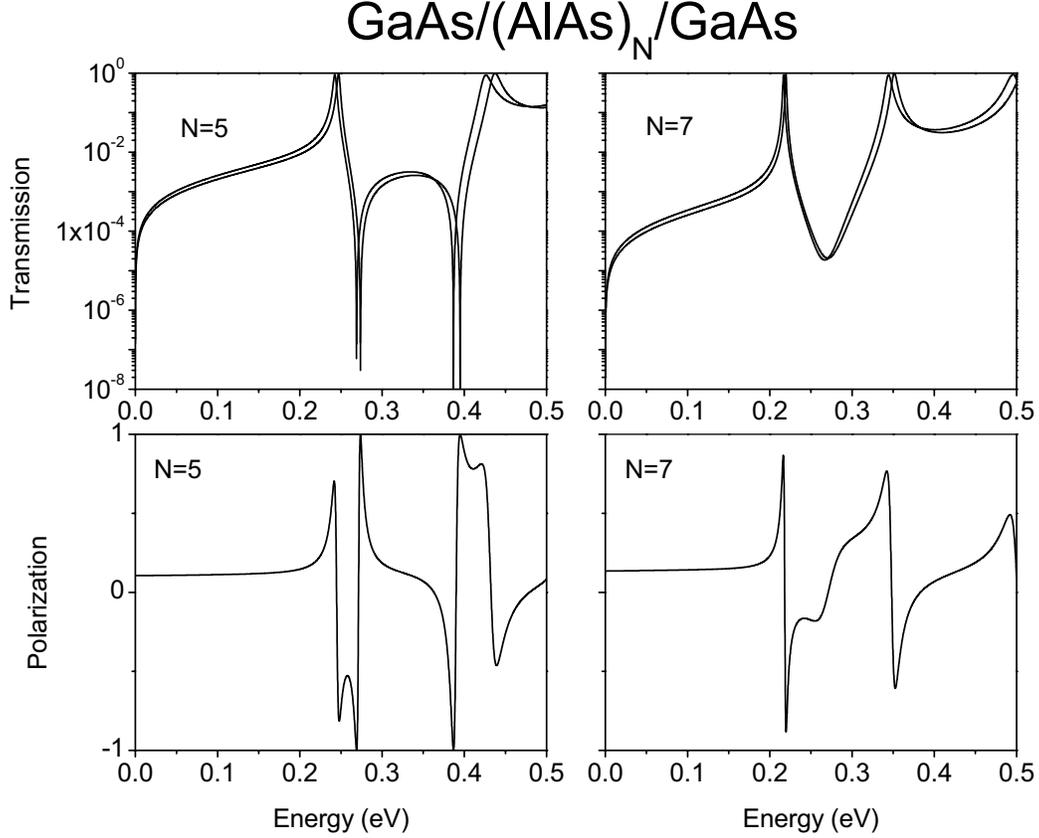}
\caption{\label{fig:5} Spin dependent transmission probability  and spin polarization of a 
GaAs/(AlAs)$_{N}$/GaAs heterostructure. 
$N$ = 5 and 7. }
\end{figure*}

In Fig.~\ref{fig:5} we plot the 
transmission probability and spin polarization for GaAs/(AlAs)$_{N}$/GaAs with $N$ = 5 and 7. 
The parity of the number of AlAs monolayers has been taken into account. \cite{Fu93} The case $N$= 5 
shows two wide windows 
with large and opposite spin polarizations. Again, the polarization windows are mainly determined by 
the resonance and 
anti-resonance positions and not by the magnitude of spin splitting. However, $N$ = 7 shows no such energy 
windows because the zeroes have moved away from 
the real axis. Moreover, at larger values of $N$ no energy windows with large polarization are found 
for both even and odd 
values of $N$. 

\begin{figure*}
\includegraphics{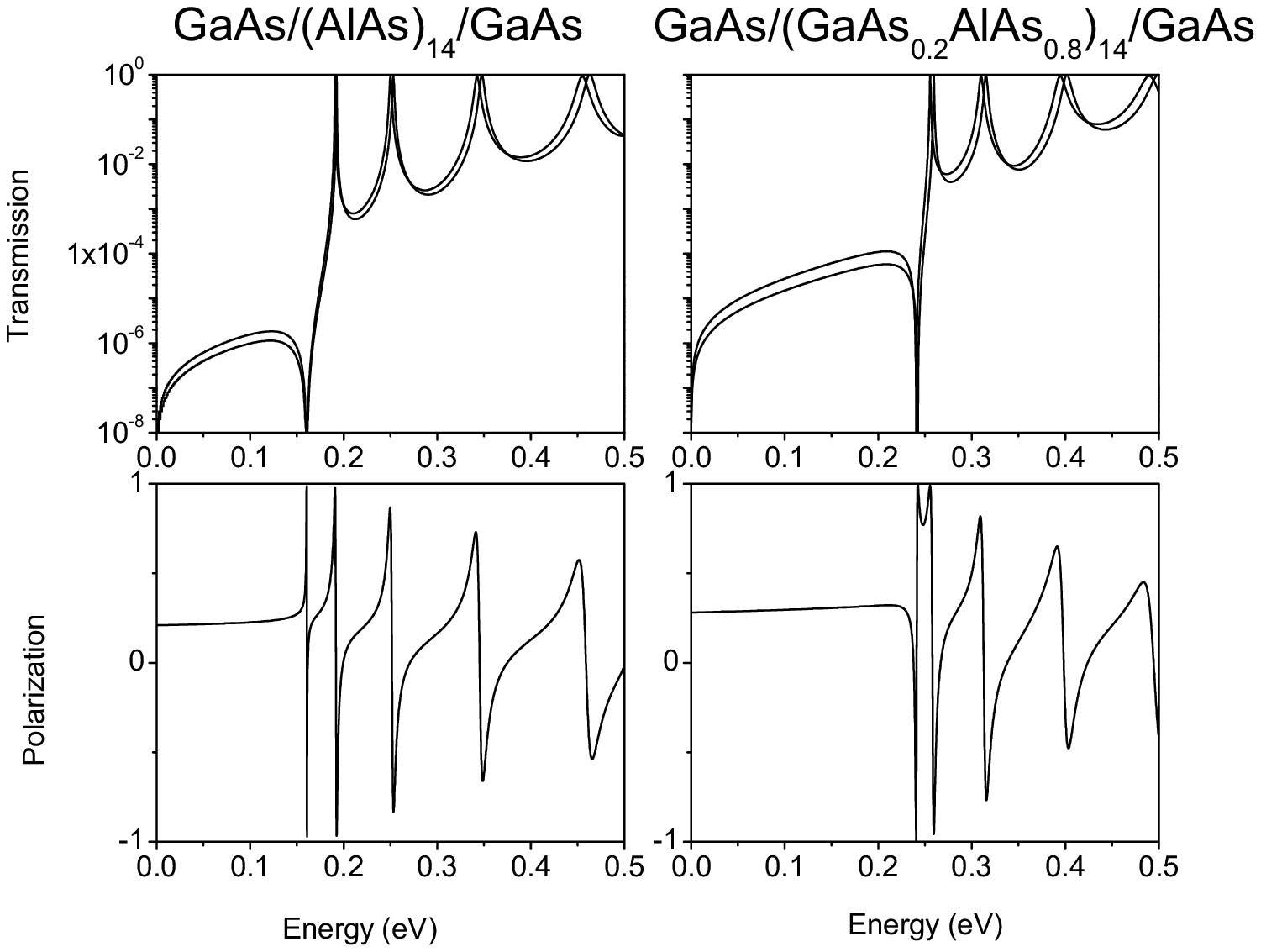}
\caption{\label{fig:6} Comparison of spin dependent transmission probability  and spin polarization between 
GaAs/(AlAs)$_{12}$/GaAs and GaAs/(Al$_{0.8}$Ga$_{0.2}$As)$_{12}$/GaAs heterostructures.}
\end{figure*}

The pattern can be re-established by increasing the strength of the Dresselhaus 
coefficient at $X$ point in the barrier. A barrier made of Al$_{0.8}$Ga$_{0.2}$As can 
achieve this goal. In Al$_{0.8}$Ga$_{0.2}$As, the 
Dresselhaus coefficients are mixtures of 
those of AlAs and GaAs. GaAs has larger coefficients, 
in particular, $\beta$ is ten times larger than the $\beta$ coefficient of AlAs, making the $\beta$ 
coefficient of the compound stronger than that 
of AlAs.  In Fig.~\ref{fig:6} we make a comparison between Al$_{0.8}$Ga$_{0.2}$As and 
AlAs barriers with a thickness of $N$=12 monolayers. Virtual crystal approximation was employed to calculate 
the physical parameters of 
Al$_{0.8}$Ga$_{0.2}$As. This is a reasonable assumption, since AlAs and GaAs have similar structural and 
electronic properties.
Fig.~\ref{fig:6} illustrates clearly that Al$_{0.8}$Ga$_{0.2}$As barrier shows a window of 
polarization, while AlAs barrier does not. 

\begin{figure*}
\includegraphics{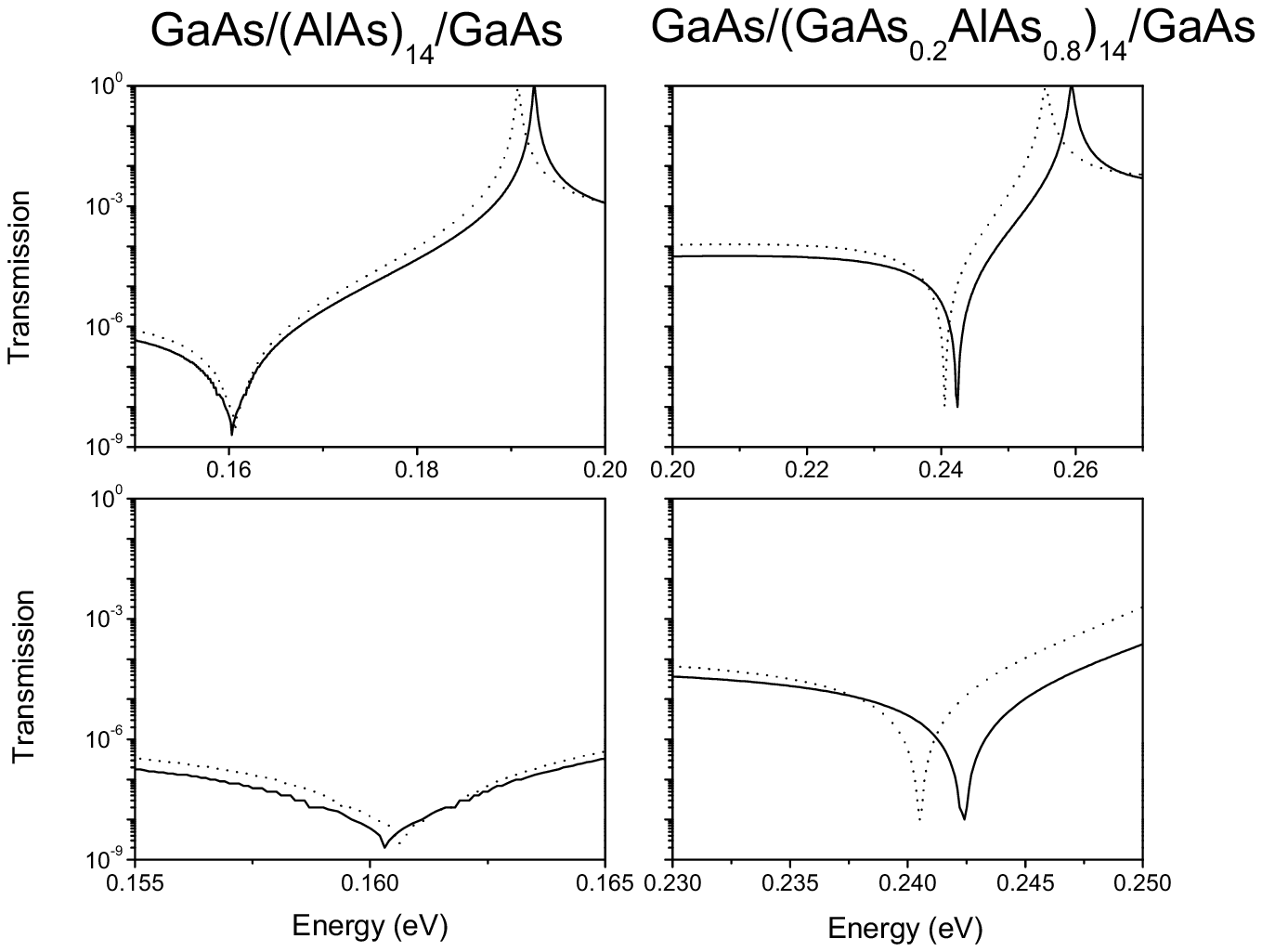}
\caption{\label{fig:7} Closer look at spin dependent transmission probability around resonance energy for 
GaAs/(AlAs)$_{12}$/GaAs and GaAs/(Al$_{0.8}$Ga$_{0.2}$As)$_{12}$/GaAs heterostructures. It explains 
the origin of energy window with 
large spin polarization.}
\end{figure*}

Visual analysis of Fig.~\ref{fig:6} and Fig.~\ref{fig:7} suggests that in order to 
obtain polarization windows, the 
resonances and zeroes in the spin channels must be properly ordered. If a resonance is occurring 
first in the spin channel 1, the zero in the transmission 
coefficient of the spin channel 1 must precede the zero in the transmission coefficient 
of spin channel 2. This condition is fulfilled for wider barriers provided that
the Dresselhaus coefficient at $X$ point is sufficiently large.

The practical aspect of focusing the electrons to energies within the large-polarization window can be 
achieved by placing an RTD structure in front of 
the tunneling barrier. In this way, one not only can control the incoming energy of the electrons but also 
ensures that most electrons have non-vanishing transverse momenta\cite{Sandu03} and, consequently, 
their energies are spin-split.

It is important to analyze at this point the influence that the neglect of the Dresselhaus $k^3$ term 
in the leads can have on the conclusions drawn in the present work.
Since the tight-binding basis is localized, different basis can be 
used to treat the 
spin Hamiltonian in barrier and in the leads. In the 
leads, the basis is that which makes diagonal the Hamiltonian  in Eq.~(\ref{eq:DB}), while in the barrier 
the basis is given 
in Eq.~(\ref{eq:EV}). The non-diagonal spin dependent part is then transferred to interface terms between 
leads and barrier. The corrections to 
the Green function and transmission coefficients are quadratic with respect 
the strength of this non-diagonal term. 
This analysis indicates that the 
conclusions of the article are not likely to change if the Dresselhaus term in the leads 
is properly taken into account. A detailed quantitative discussion 
will be presented elsewhere since special care must be exercised due to the
presence of the $k_z$-linear term in Eq.~(\ref{eq:DB}).

\section{Conclusions}

We investigated the spin dependent transport across an 
indirect semiconductor barrier of a zinc-blende structure like GaAs/AlAs/GaAs heterostructure along 
[001] axis by means of a combination of several tight-binding models. 
Spin tunneling through such an indirect barrier exhibits two major characteristics: the 
proportionality of the Dresselhaus Hamiltonians at $\Gamma$ and $X$ points and the Fano resonances.
A generic tight-binding Hamiltonian has shown that large spin polarization occurs in the energy window 
determined by the separation between the resonance and its associated anti-resonance and not 
by the magnitude of the spin splitting of resonances. Moreover, the energy separation 
between the resonance and its corresponding anti-resonance increases as the barrier width increases.

Realistic calculations have been performed with a two-band tight-binding model. The  effective mass 
Hamiltonians at  $\Gamma$ and $X$ have been converted\cite{Frensley94} to the tight-binding Hamiltonians. 
The $\Gamma$-$X$ coupling was implemented following the 
scheme presented in Ref.~\onlinecite{Fu93}. Accordingly, the Dresselhaus Hamiltonians at $\Gamma$ and $X$ 
in the barrier have been included in the effective masses and band offsets. The calculations show that, in order 
to obtain energy windows with large polarization, two conditions need to be satisfied. 
The first condition consists of having well separated 
resonances such that their corresponding anti-resonances do not interact with 
each other. The second condition is that the relative energy order of the 
resonances in the two spin channels must be the same as
the order of their corresponding zeroes.
The first condition is 
achieved by an appropriate combination of barrier width and longitudinal effective mass at $X$ point, 
while the second condition is accomplished by a combination of spin splitting strength at $X$ point 
and transmissibility through the direct barrier. 

Electrons can be focused in the required energy windows by passing them 
through a resonant tunneling diode structure situated in front of the indirect barrier. 
Using such an experimental setup, one could obtain large spin polarization following the procedure of 
Perel {\it et al.}\cite{Perel03} and Glazov {\it et al.}\cite{Glazov05}.

\begin{acknowledgments}
This work has been supported in part by NSERC grants no. 311791-05 and 315160-05.
The authors wish to acknowledge generous support in the form of computer resources from
the R\'{e}seau Qu\'{e}b\'{e}cois de Calcul de Haute Performance.
\end{acknowledgments}




\end{document}